\RequirePackage{fix-cm}
\documentclass[smallextended]{svjour3}       
\smartqed  
\usepackage{graphicx}
 \journalname{Journal of Transportation Security}
\begin{document}

\title{A tale of two cities
}

\subtitle{Vulnerabilities of the London and Paris transit networks}


\author{Christian von Ferber         \and
        Bertrand Berche \and Taras Holovatch \and Yurij Holovatch
}


\institute{Christian von Ferber \at
              Applied Math. Research Centre, Coventry University, UK
           \and
           Bertrand Berche \at
              Institut Jean Lamour, Universit\'e de Lorraine,
              Vand\oe uvre les Nancy, France
              \and
Taras Holovatch \at
              Applied Math. Research Centre, Coventry University, UK \&  Institut Jean Lamour, Universit\'e de Lorraine,
              Vand\oe uvre les Nancy, France
              \and
Yurij Holovatch \at Institute for Condensed Matter Physics, National
Acad. Sci. of Ukraine, Lviv, Ukraine
}

\date{Received: date / Accepted: date}

\maketitle

\begin{abstract}
This paper analyses the impact of random failure or attack on the
public transit networks of London and Paris in a comparative study.
In particular we analyze how the dysfunction or removal of sets of
stations or links (rails, roads, etc.) affects the connectivity
properties within these networks. We show how accumulating
dysfunction leads to emergent phenomena  that cause the
transportation system to break down as a whole. Simulating different
directed attack strategies, we find minimal strategies with high
impact and identify a-priory criteria that correlate with the
resilience of these networks. To demonstrate our approach, we choose
the London and Paris public transit networks. Our quantitative
analysis is performed in the frames of the complex network theory --
a methodological tool that has emerged recently as an
interdisciplinary approach joining methods and concepts of the
theory of random graphs, percolation, and statistical physics. In
conclusion we demonstrate that taking into account cascading effects
the network integrity is controlled for both networks by less than
0.5\% of the stations i.e. 19 for Paris and 34 for London.
\end{abstract}

\section{Introduction}
\label{I}

 "The traveller fared slowly on his way, who fared towards Paris from England in the
autumn of the year one thousand seven hundred and ninety-two. More
than enough of bad roads, bad equipages, and bad horses, he would
have encountered to delay him".\footnote{Charles Dickens. {\em A
Tale of Two Cities}. London: Penguin Classics (2003).} In the times
when Charles Dickens wrote these words in his famous novel, there
was perhaps not too much difference concerning the quality of
transportation systems in the two cities, or reasons that may have
caused their malfunction. Historical circumstances may have had a
different impact on the development of public transit security in
both cities. However, today one might assume that on average the
differences observed between the facilities offered by
transportation networks of developed countries may be small enough.
Analyzing the readily available data on these networks both with
algorithms and analytical approaches we test this assumption and
quantify remaining differences.

The aim of this paper is to compare security features of highly
developed contemporary public transit networks (PTN) - of two
European capitals, London and Paris. These cities were chosen as
they display similarities in their structure and historical
development caused by geographical and social reasons and further
due to the particular role of the public transit facilities of
London in the wake of the 2012 Olympics. We will be interested in
the impact of both random failure and targeted attacks that may lead
to dysfunction either within the stations of the PTN or along the
links (rails, roads, bridges, etc.) that connect them. In a general approach
to this problem one would want to consider a dynamic model of the PTN
including the current local capacities and loads at the time of failure,
detailed passenger destinations and itinerary together with a full view
of the PTN structure (i.e. topological  and connectivity properties of
the network). In lack of availability of corresponding data for such an
approach we restrict our study to the impact of failure on the topological
and connectivity properties of the analyzed networks.
This provides a first but essential step towards understanding how vulnerabilities may
be reduced by choosing appropriate network topologies.
In particular, we will consider the static network structure of the PTNs of
London and Paris and analyze their vulnerability with respect to dysfunction due
to random failure or directed attack.
As we will see below, simulating various failure and attack scenarios even on
this level illuminates significant differences and allows for general conclusions
concerning the behavior of these PTNs under stress.

The setup of this paper is as follows. In the next section we briefly describe the general
method of our analysis -- complex network theory
\cite{network_reviews,network_books} -- and present an overview on previous
studies. We proceed to discuss the problem of PTN vulnerability in section
\ref{III}, where we show how this problem is related to the percolation
theory \cite{percolation}.
We introduce observables that quantitatively measure the impact on PTNs under
attack, a problem we further analyze in section \ref{IV} where we present a comparative analysis
of the London and Paris PTNs and the impact of failure and attacks of different nature.
We discuss possible reasons for the differences observed for PTN vulnerability and propose
estimators for local and global properties that allow a priory assessment of the degree
of resilience or vulnerability of PTNs. Taking into account cascading effects in the interplay
between routes and stations we demonstrate in section \ref{V} that the network integrity
hinges on the effective operation of a very small set of important stations.

\section{A complex network model of public transit}
\label{II}

The observation that the paths of public transit routes of a city
form a network and that this network is complex enough is  part of
our everyday experience. However, the concept of {\em complex
networks} has recently become the nucleus of a new and rapidly
developing field of knowledge that has its roots in random graph
theory and statistical physics (see e.g. recent reviews
\cite{network_reviews} and monographs \cite{network_books}). From a
mathematical point of view, a network is nothing else but a graph
defined by a set of vertexes and a set of edges or links each
connecting a pair of vertexes. Graph theory is well-settled branch
of discrete mathematics with origins in classical works of L. Euler
\cite{book_graph}. An essential breakthrough and a paradigm shift in
graph theory (and in particular concerning random graphs) occurred
in the 1990-ies, when particular correlations were discovered in
other\-wise seemingly random graphs. It was realized, that numerous
natural and man-made structures may be described in terms of
networks and that these networks posses surprising properties,
strikingly different from those of the so-called classical random
graph \cite{Bollobas85}. Such networks are currently classified as
complex networks. To name a few, these include networks describing
interac\-ting systems of biological, ecological, sociological or
technological origins such as networks of cell metabolism,
communication, transportation, and many other forms of interaction.
Complex networks have been found to be compact structures (sometimes
called {\em small worlds}) with short distances between nodes, and a
high level of correlation and self-organization. They demonstrate
extremely high resilience with respect to random failure. However
they are proven to be particularly vulnerable with respect to
targeted attacks. Some of their statistical properties, in
particular the distribution of node degrees (the number of
connections of individual nodes) are governed by power laws. This
indicates the presence of non-trivial correlations within the
structure of these systems. We set out to show that similar
properties are inherent to the PTN of London and Paris studied in
the present work.

Recent research
\cite{Latora,Seaton04,metro,Holyst,china,vonFerber,Derrible11} on
public transit networks has produced quantitative evidence that PTNs
share general features of other transportation or transmission
networks like airport, railway, or power grid networks
\cite{network_reviews}. These features include evolutionary growth,
optimization, and usually an embedding in two dimensional (2D)
space. Earlier empirical studies of PTNs in the frames of complex
network theory have often restricted the analysis to certain
sub-networks of city transit. Examples are studies of subway
networks of Boston \cite{Latora,Seaton04}, Vienna \cite{Seaton04}
and several other cities \cite{metro}, and city bus networks in
Poland \cite{Holyst} and China \cite{china}. However, as far as the
bus-, subway- or tram-subnetworks are not closed systems the
inclusion of additional subnetworks has significant impact on the
overall network properties as has been shown for the subway and bus
networks of Boston \cite{Latora}. Therefore, further analysis of PTN
has included the full set of subnetworks \cite{vonFerber}.

\begin{table}
\caption{ Characteristics of the PTNs analyzed in this study. $N$:
number of stations; $R$: number of routes. Given characteristics
are: $\langle k \rangle$ (mean node degree); $\ell^{\rm max}$,
$\langle \ell \rangle$ (maximal and mean shortest path length);
${\cal C}$ (relation of the mean clustering coefficient to that of
the classical random graph of equal size, (\ref{3}) ); ${\cal C}_B$:
betweenness centrality (\ref{5}); $\kappa^{(z)}$, $\kappa^{(k)}$
(c.f. Eqs. (\ref{15}), (\ref{13})); degree distribution exponent
$\gamma$ (\ref{4}). Additional details may be found in
\cite{vonFerber}.}
\label{tab1}       
\begin{tabular}{lllllllllll}
\hline\noalign{\smallskip} City &   $N$ & $R$ & $\langle k \rangle$
& $\ell^{\rm max}$ & $\langle \ell \rangle$ & ${\cal C}$ & ${\cal C}_B$ & $\kappa^{(z)}$ &
$\kappa^{(k)}$ & $\gamma$ \\
\noalign{\smallskip}\hline\noalign{\smallskip}
London &  10937 &   922 &  2.60  &  107 &  26.5 &  320.6  & 1.4$\cdot 10^5$ & 1.87 & 3.22 & 4.48 \\
Paris  &  3728  &   251 &  3.73  &  28  &   6.4 &  78.5   & 1.0$\cdot 10^4$ & 5.32 & 6.93 & 2.62 \\
\noalign{\smallskip}\hline
\end{tabular}
\end{table}

The two PTNs analyzed within the present work are either operated by
a single operator (Traffic for London, TFL) or by small number of
operators with a coordinated schedule (three operators for Paris),
as expressed by a central web site from which our data was
obtained.\footnote{See \cite{vonFerber} for more details on the
database.} The analyzed PTN of London covers the metropolitan area
of 'Greater London' and includes buses, subway, and tram.
Correspondingly, the PTN of Paris as analyzed here covers the
metropolitan area 'aire urbaine' and comprises buses, RER and
subway. Some characteristics of these networks are given in Table
\ref{tab1}. There is a number of different ways to represent a PTN
in the form of a graph
\cite{Latora,Seaton04,metro,Holyst,china,vonFerber,Derrible11,Holovatch11}.
In what follows below, we will mostly use the so-called ${\bf
L}$-space representation \cite{Latora,Holyst,vonFerber}, where each
public transit station is represented by a vertex (node) and any two
stations serviced successively by at least one route are connected
by an edge (link). In this representation the obtained graph -- a
complex network -- is most similar to the PTN map.\footnote{Note,
however that multiple links are absent in this graph.} Typical
measures for the 'diameter' of the network are the maximal or the
mean {\em shortest path} lengths $\ell^{\rm max}$ and $\langle \ell
\rangle$. The latter is defined by:
\begin{equation}\label{1}
\langle \ell \rangle = \frac{2}{N(N-1)}\sum_{i>j \in \cal{N} }\ell
(i,j),
\end{equation}
where $N$ is the number of network nodes, $\ell (i,j)$ is the length
of a shortest path (in terms of station intervals traveled) between
nodes $i$ and $j$ and the sum spans all pairs $i,j$ of sites that
belong to the network $\cal{N}$. The comparatively low values of
$\langle \ell \rangle$ found for the two PTNs (see Table \ref{tab1})
may be related to their small world structure (where
$\langle\ell\rangle$ shows logarithmic growth with the number of
nodes) \cite{vonFerber}. The fact that the London PTN has a larger
value $\ell^{\rm max}$ corresponds to the larger area covered by the
network (as seen, e.g. from larger number of routes and stations).

The mean and maximal shortest path lengths characterize the network
as a whole and sometimes are referred to as global properties of the
network.  An example of a local property is given by the node degree
$k_i$, the number of links that are connected to the node $i$. By
definition, it is equal to the number of nodes adjacent to the given
one and defines the neighborhood size of this node. Obviously, not
all neighbors of the node $i$ need to be neighbors of each other.
This property is measured by the {\em clustering coefficient}:
\begin{equation}\label{2}
C_i= \frac{2y_i}{k_i(k_i-1)}, \hspace{1em} k_i\geq 2,
\end{equation}
where $y_i$ is the number of links between the neighbors of node $i$
and $C_i = 0$ for $k_i = 0, 1$. In general, clustering reflects a
specific form of correlation present in a network:  the clustering
coefficient of a node may also be interpreted as the probability of
any two of its neighbors to be connected. A useful numerical
indicator is given by the ratio of the mean clustering coefficient
of a network to the corresponding value for the classical
Erd\"os-R\'enyi random graph of equal size:
\begin{equation}\label{3}
{\cal C} = \langle C_i \rangle/C_{ER}.
\end{equation}
Here, $C_{ER}=2M/N^2$ where the classical Erd\"os-R\'enyi random
graph is constructed by randomly linking $N$ nodes by $M$ links
\cite{network_reviews,network_books}. Therefore, the high values of
${\cal C}$ found in Table  \ref{tab1} for London and Paris indicate
strong local correlations in these networks. Moreover, the London
PTN appears locally to be stronger correlated than that of Paris.

Another striking difference between the properties of random graphs
and the PTNs considered here is the behavior of the {\em node-degree
distribution} $P(k)$, the probability that an arbitrary node is of
degree $k$. The random graph is characterized by a Poisson
distribution which decays exponentially for large $k$
\cite{network_reviews,network_books}. The empirically observed
distributions for the London and Paris PTNs however show a decay
best described by a power law \cite{vonFerber} :
\begin{equation}\label{4}
P(k) \sim k^{-\gamma}, \hspace{1em} k\gg 1.
\end{equation}
The power law decay (\ref{4}) indicates {\em scale-free} properties
of the London and Paris  PTNs. It is instructive to note that the
exponent $\gamma$ governing this decay is much smaller for the PTN
of Paris, see Table \ref{tab1}. As we will show this has important
impact on the observed resilience of the network.

To some extent, the node degree may be considered as a local measure
of the importance of the node: it is intuitively reasonable that
hubs (nodes with a high degree) play an essential role in networks.
The importance of a node with respect to the connectivity between
other nodes of the network, however, is more efficiently measured by
the so-called {\em betweenness centrality}. For a given node $i$,
the latter measures the overall share of shortest paths between
pairs of other nodes that pass through this node. The betweenness of
node $i$ may be calculated as:
\begin{equation}\label{5}
{\cal C}_B (i) = \sum_{j\neq i \neq k \in \cal{N}}
\frac{\sigma_{jk}(i)}{\sigma_{jk}}
\end{equation}
where $\sigma_{jk}$ is the number of shortest paths between nodes
$j$ and $k$ of the network $\cal{N}$ and $\sigma_{jk}(i)$ is the
number of these paths that go via node $i$. Numerical values of the
mean betweenness are given in Table \ref{tab1} for both PTNs.

\section{PTN resilience: observables and attack scenarios}
\label{III} The impact on complex network behavior upon removal of
either their nodes or links is closely related to so-called lattice
percolation phenomena \cite{percolation}. The latter occurs on
homogeneous structures (lattices) whereas the non-homogeneity of
complex networks gives rise to a variety of phenomena specific for
these structures. The empirical analysis of  scale-free real-world
networks has shown that these networks display unexpectedly high
degrees of robustness under random failure
\cite{network_reviews,network_books}. However they may be
particularly vulnerable to attacks, that target important nodes or
links. As we have seen in the previous section, both the London and
the Paris PTN share common features of complex networks. Therefore,
we may expect their behavior under stress or attack to be similar.

The first property a transit network trivially needs to fulfil is
overall connectivity: there must be a path within the network
between any two nodes. Upon failure of a smaller or larger set of
nodes this overall connectivity may get lost. Generally one
considers a network to remain functional if a significantly large
connected component (sometimes called a spanning cluster) remains
connected.

The phenomenon of the appearance and nature of such spanning
clusters is at the center of a well established field of Statistical
Physics: {\em percolation theory} \cite{percolation}. Originally it
describes the emergence of such spanning clusters on a lattice at a
certain threshold for the concentration $c_{\rm perc}$ of links or
nodes present on the lattice and predicts universal properties that
may be observed and calculated within the theory with high
precision.

On a lattice, the appearance of a spanning cluster signals the onset
of percolation at a particular concentration $c_{\rm perc}$ of
lattice occupation. In turn, the probability that an arbitrary
chosen lattice site belongs to the spanning cluster is naturally
used as an order parameter: it is equal one for $c=1$, zero for $c <
c_{\rm perc}$ and follows universal behavior as $c$ approaches
$c_{\rm perc}$ from above. A similar percolation phenomenon occurs
when a {\em giant connected component} emerges on an idealized
complex network. The giant connected component  is understood as a
connected subnetwork which in the limit of an infinite network
contains a finite fraction of the network. As far as real world
networks are finite, the giant component is not well defined.
Instead we will observe the size $N_1(c) $ of the largest connected
component in the network to monitor the behavior of the network as
function of the share $c$ of nodes or links that are removed in
sequence. For convenience we define the relative size (or share) of
the largest component as the ratio of $N_1(c)$ to $N$, the number of
nodes in the initial unperturbed network:
\begin{equation}\label{6}
S(c)=N_1(c)/N.
\end{equation}
Another variable that may be used to monitor changes in network
structure as nodes or links are removed is the mean inverse shortest
path length \cite{Holme02}:
\begin{equation}\label{7}
\langle \ell^{-1} \rangle = \frac{2}{N(N-1)}\sum_{i>j \in \cal{N}
}\ell^{-1} (i,j).
\end{equation}
Here, as in (\ref{1}), $\ell (i,j)$ is the shortest path between
nodes $i$ and $j$ that belong to the network $\cal{N}$. Note
however, that while (\ref{1}) is ill-defined for the disconnected
network, the quantity (\ref{7}) is well-defined as far as $\ell^{-1}
(i,j)=0$ if nodes $i,j$ are disconnected. It may therefore be used
to trace impact on the network under attack.

\begin{figure}
\includegraphics[height=0.22\textwidth]{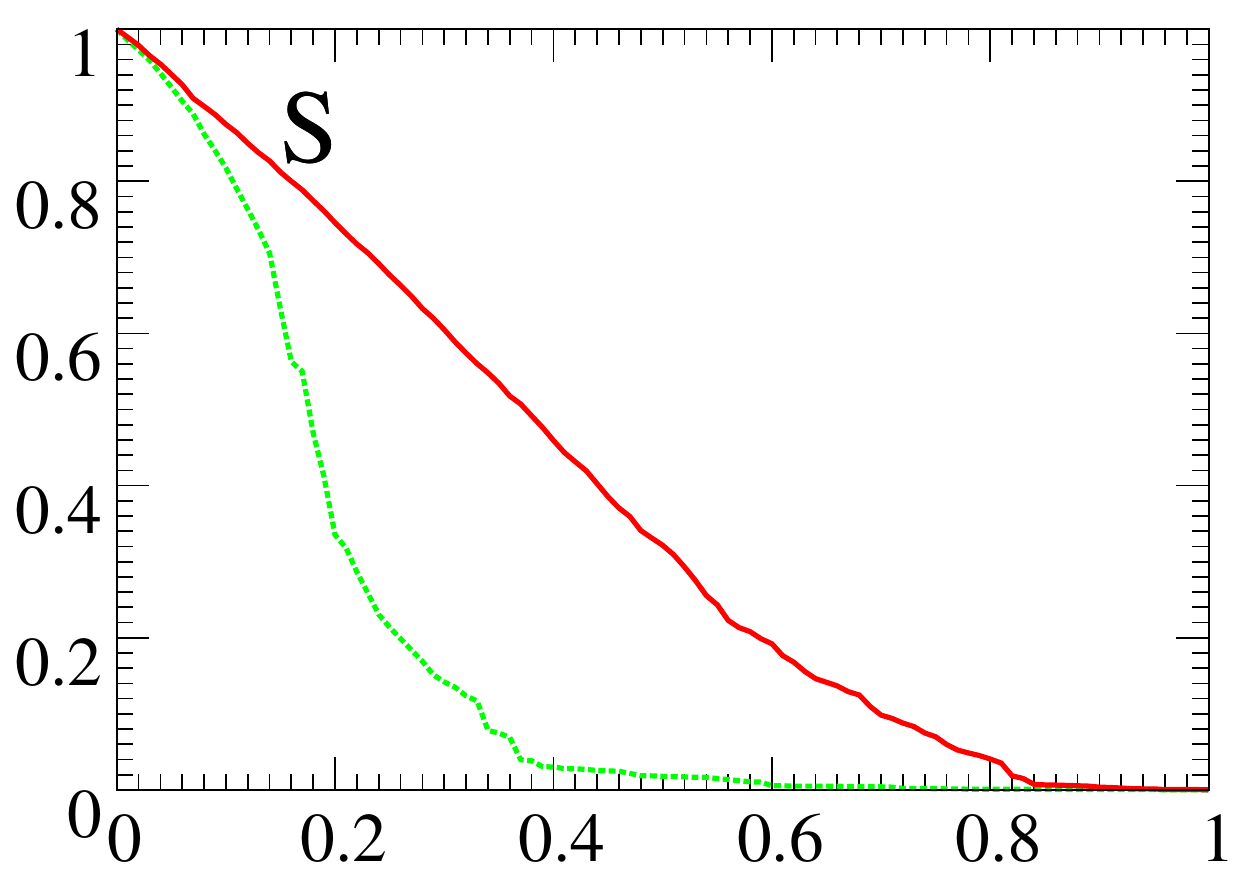}
    \includegraphics[height=0.22\textwidth]{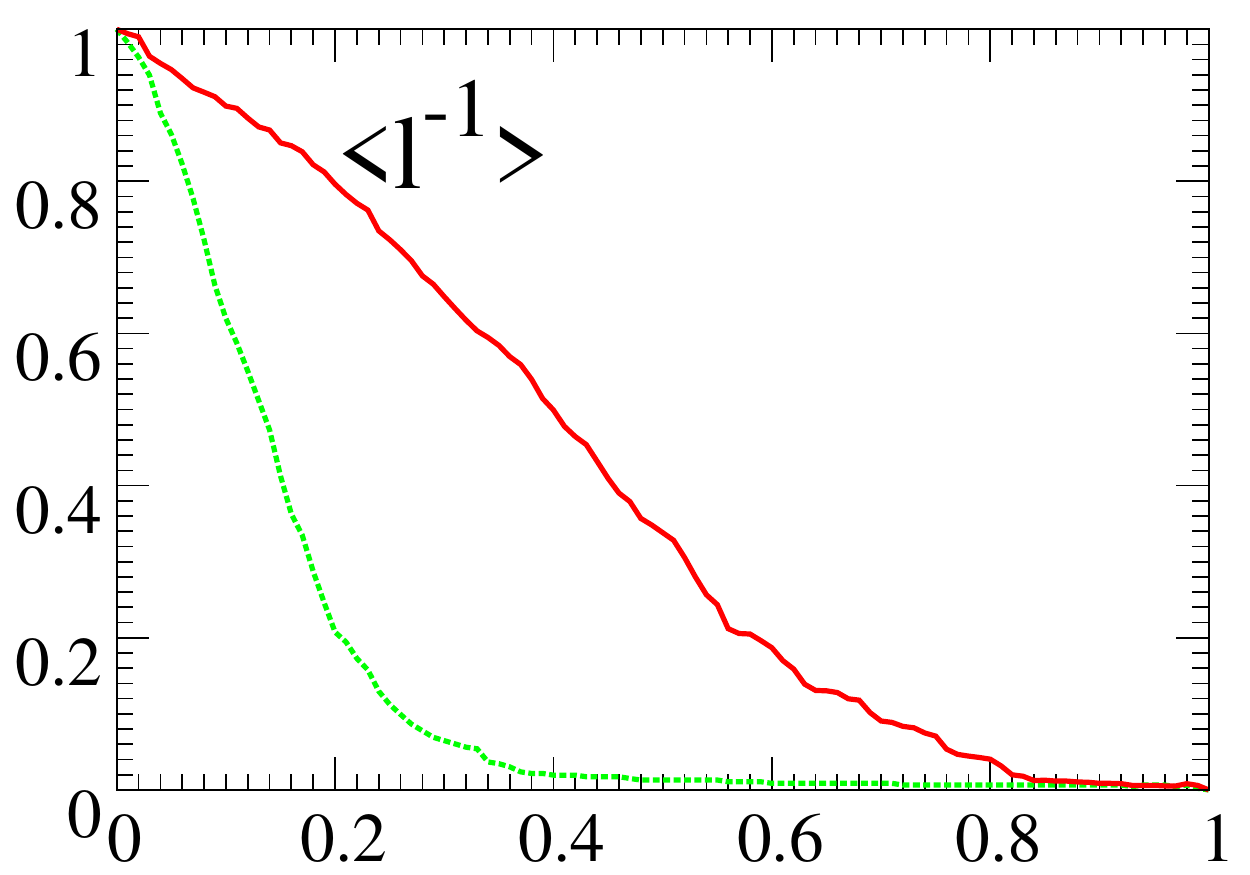}
\includegraphics[height=0.22\textwidth]{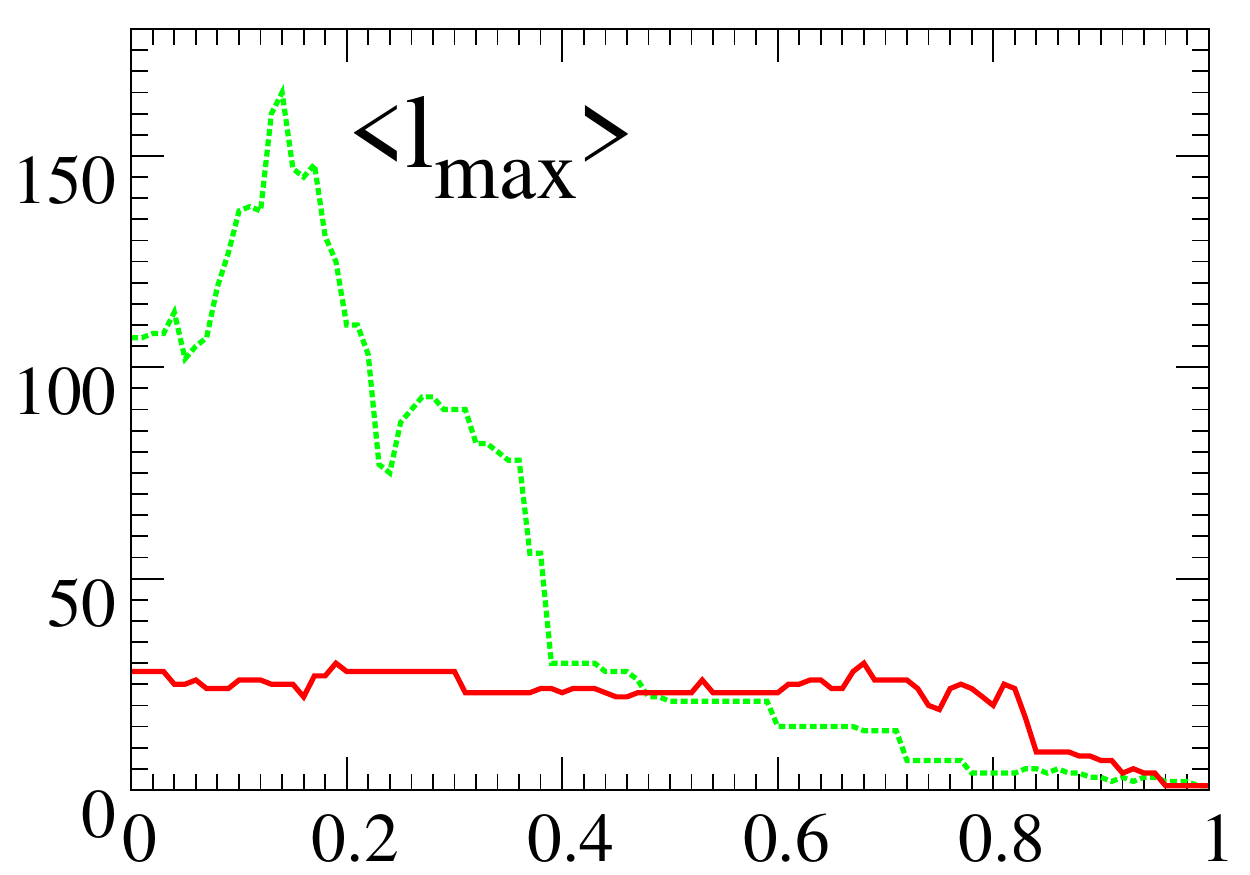}\\
\centerline{\bf a. \hspace{10em} b. \hspace{10em} c.}
\caption{\label{fig1}Share of the largest component $S$ ({\bf a}),
mean inverse $\langle \ell^{-1} \rangle$ ({\bf b}) and maximal
$\ell^{\rm max}$ ({\bf c}) shortest path length as function of the
removed share $c$ of nodes for the PTN of London (light green curve)
and Paris (dark red curve). Random removal of PTN nodes.}
\end{figure}

In Figs. \ref{fig1} {\bf a}, {\bf b} we show the behavior of $S$ and
$\langle \ell^{-1} \rangle$ for the PTN of London and Paris  as
function of the share of removed nodes $c$, as these are removed at
random. Already this simple random approach to probe the PTN
behavior under attack shows a higher vulnerability of the London PTN
to random removal of its nodes: both the $S$- and $\langle \ell^{-1}
\rangle$-curves indicate a faster decay in the case of London PTN.
Moreover, the $S$-curve for the Paris PTN decays almost linearly.
This indicates that sub-clusters less connected to the overall
network are almost absent. The size of the largest component
decreases only due to the removed nodes. This observation will be
further  quantified in the next section. Here, we want to support it
by displaying the maximal shortest path length behavior, Fig.
\ref{fig1} {\bf c}. As a matter of fact, $\ell^{\rm max}$ manifests
very different behavior for these two PTN. For the London PTN,
$\ell^{\rm max}$ grows initially and then, when a certain threshold
is reached ($c\sim 0.14$) it abruptly decreases. Obviously, removing
the nodes initially increases the path lengths as deviations from
the original shortest paths need to be taken into account. At some
point, removing further nodes then leads to a breakup of the network
into smaller components on which the paths are naturally limited
which explains the sudden decrease of their lengths. Such
peculiarities in the behavior of $\ell^{\rm max}$ are almost absent
for the Paris PTN, at least for small and medium values of $c$.

Note, that the plots of Fig. \ref{fig1} display the results of a
single random sequence of node removal. However, as we have checked
statistics over large number of random attack sequences
\cite{Holovatch11,attacks}, the large PTN size leads to a
'self-averaging'  effect: averaging over many random attack
sequences gives results almost identical to those presented in Fig.
\ref{fig1}. To further analyze the PTN attack vulnerability, we have
made a series of computer simulations removing PTN nodes and links
not at random, but ordering them according to their importance with
respect to network connectivity, scenarios we call {\em attack} or
{\em directed attack}. To order these we use the already mentioned
properties such as node degree, betweenness centrality (\ref{5}),
clustering coefficient (\ref{2}) and several other indicators (see
\cite{Holovatch11,attacks}). Another attack scenario that has proven
successful in immunization problems on complex networks
\cite{Cohen03} consists in removing of a randomly chosen neighbor of
a randomly chosen node. Its efficiency is based on the fact, that in
this way nodes with a high number of neighbors will be selected with
higher probability. Each of the above described scenarios (except
for the random ones) was realized for the lists prepared for the
initial network or lists rebuilt by recalculating the order of the
remaining nodes after each step. The latter way is known to be
usually more efficient and leads to slightly different results
suggesting that the network structure changes in the course of the
attack \cite{Holme02,Girvan02}.

\begin{figure}[h]
\includegraphics[width=0.75\textwidth]{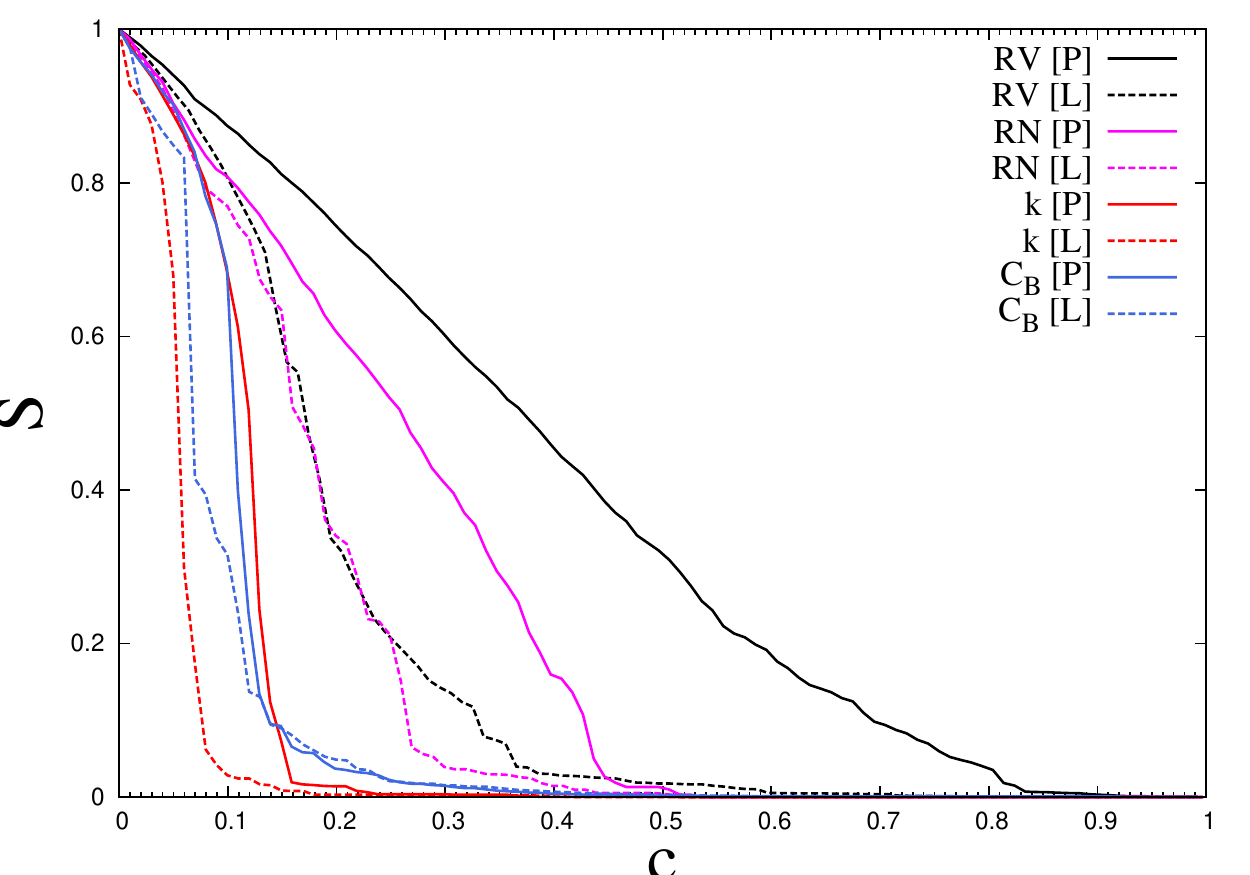}
\caption{\label{fig2}Relative size $S$ of the largest component of
the London and Paris PTNs as function of the share $c$ of removed
nodes either chosen at random, or ordered by decreasing node degree
$k$ or betweenness ${\cal C}_B$ centrality. The lists were rebuilt
by recalculating the order of the remaining nodes after each step.
RV (RN): random removal of a node (or of its randomly chosen
neighbor). A letter in square brackets refers to the London [L] or
Paris [P] PTN.}
\end{figure}

In Fig. \ref{fig2} we show the relative size of the largest
component of London and Paris PTN as function of the share of nodes
removed following specific attack scenarios described above. More
specifically, nodes were removed in chunks of 1 \% of the initial
nodes and a recalculation took place after the removal of each 1\%
chunk of nodes. As may be drawn from a first glance of the plots,
the most harmful are attacks targeted on the nodes of highest node
degree and highest betweenness. We will discuss these in more detail
in the next section, complementing the picture of node-targeted
attacks by that of attacks that target PTN links.

\section{PTN vulnerability: quantitative analysis}
\label{IV}

In what follows we discuss in some detail those attacks that have
highest impact on the two PTNs and compare these with the random
attack scenario. To this end, we introduce indicators that quantify
PTN resilience \cite{Schneider,Berche12}. Furthermore, we seek
correlations between PTN resilience and network characteristics that
may be measured independently. We apply this scheme to both
node-targeted attacks (section \ref{IV.1}) and link-targeted attacks
(section \ref{IV.2}).

\subsection{Node-targeted attacks} \label{IV.1}

\begin{figure}
\includegraphics[width=0.32\textwidth]{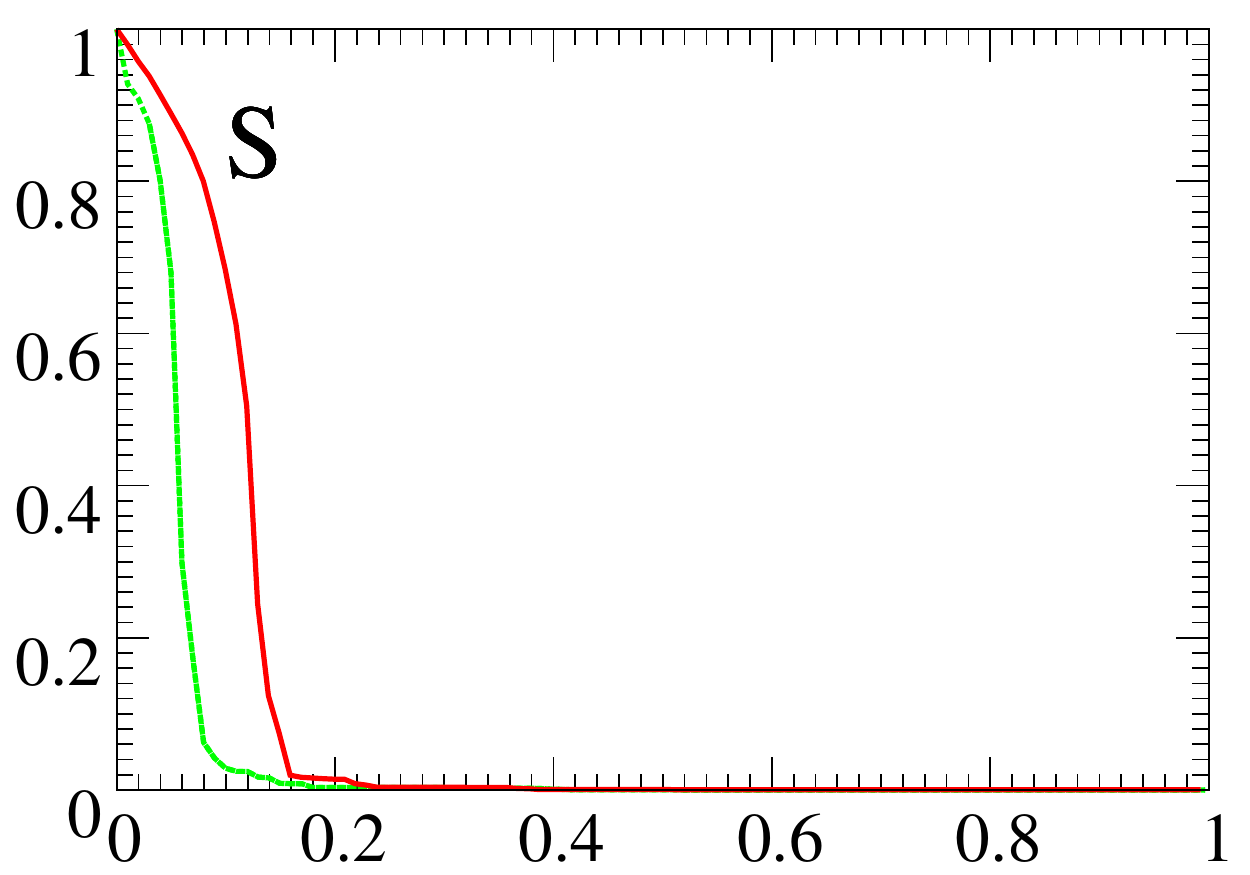}
    \includegraphics[width=0.32\textwidth]{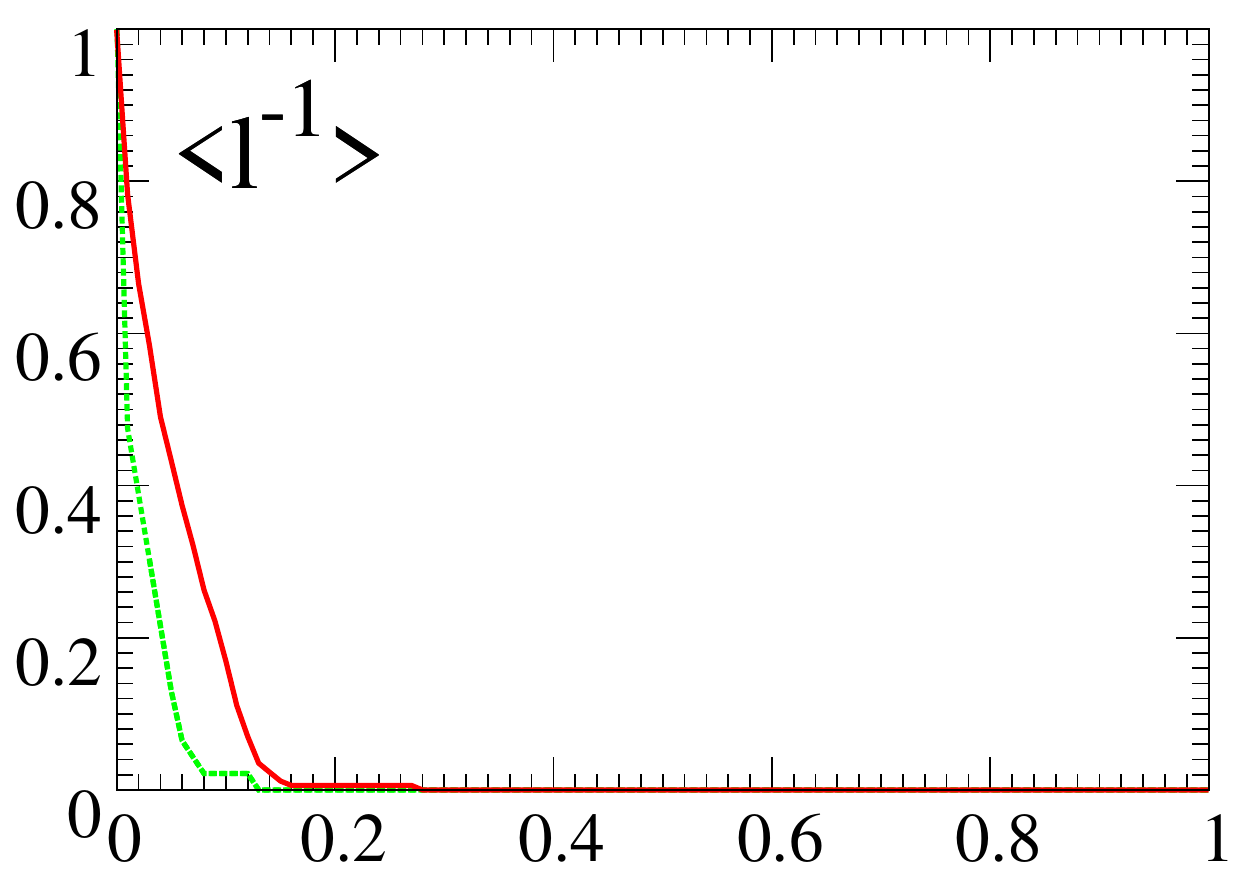}
\includegraphics[width=0.32\textwidth]{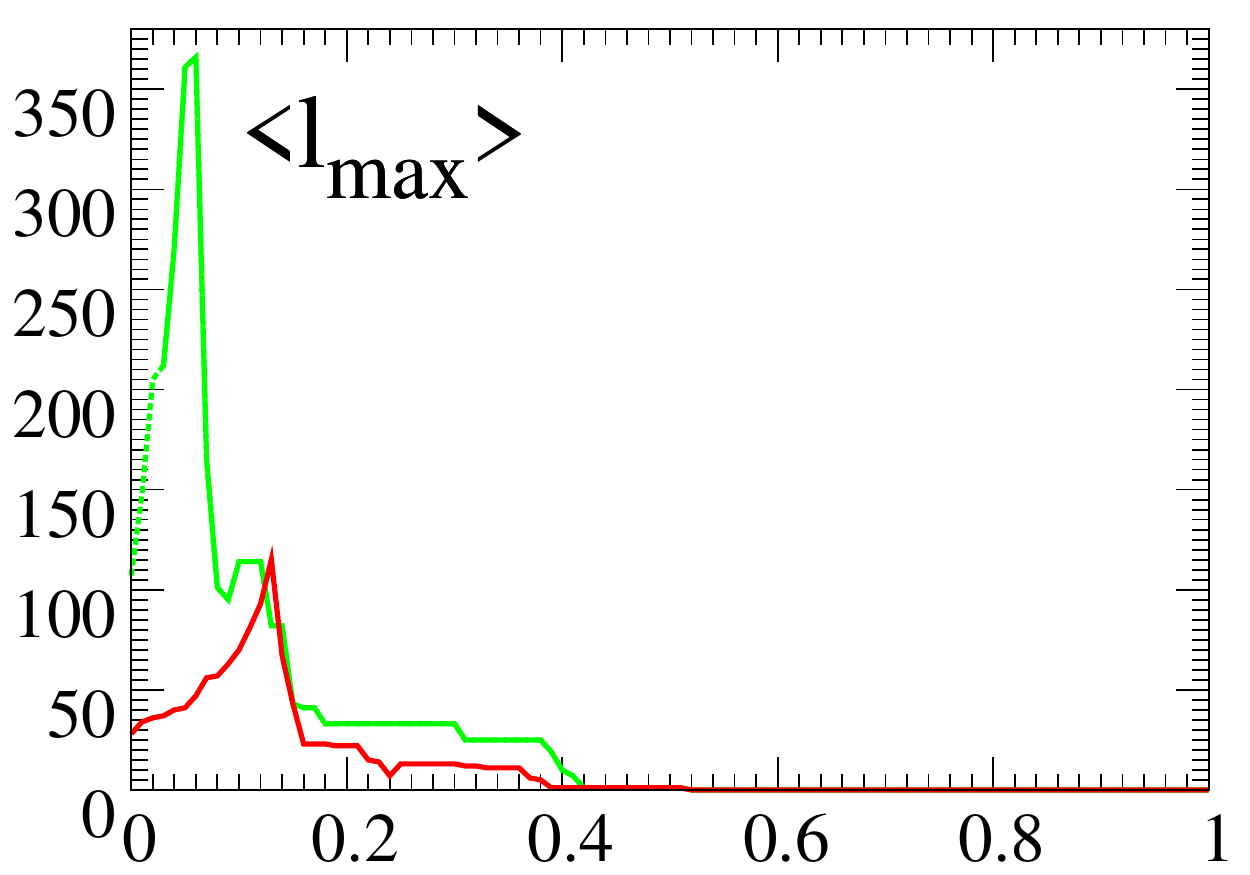}
\centerline{\bf a. \hspace{10em} b. \hspace{10em} c.} \\
\\

\includegraphics[width=0.32\textwidth]{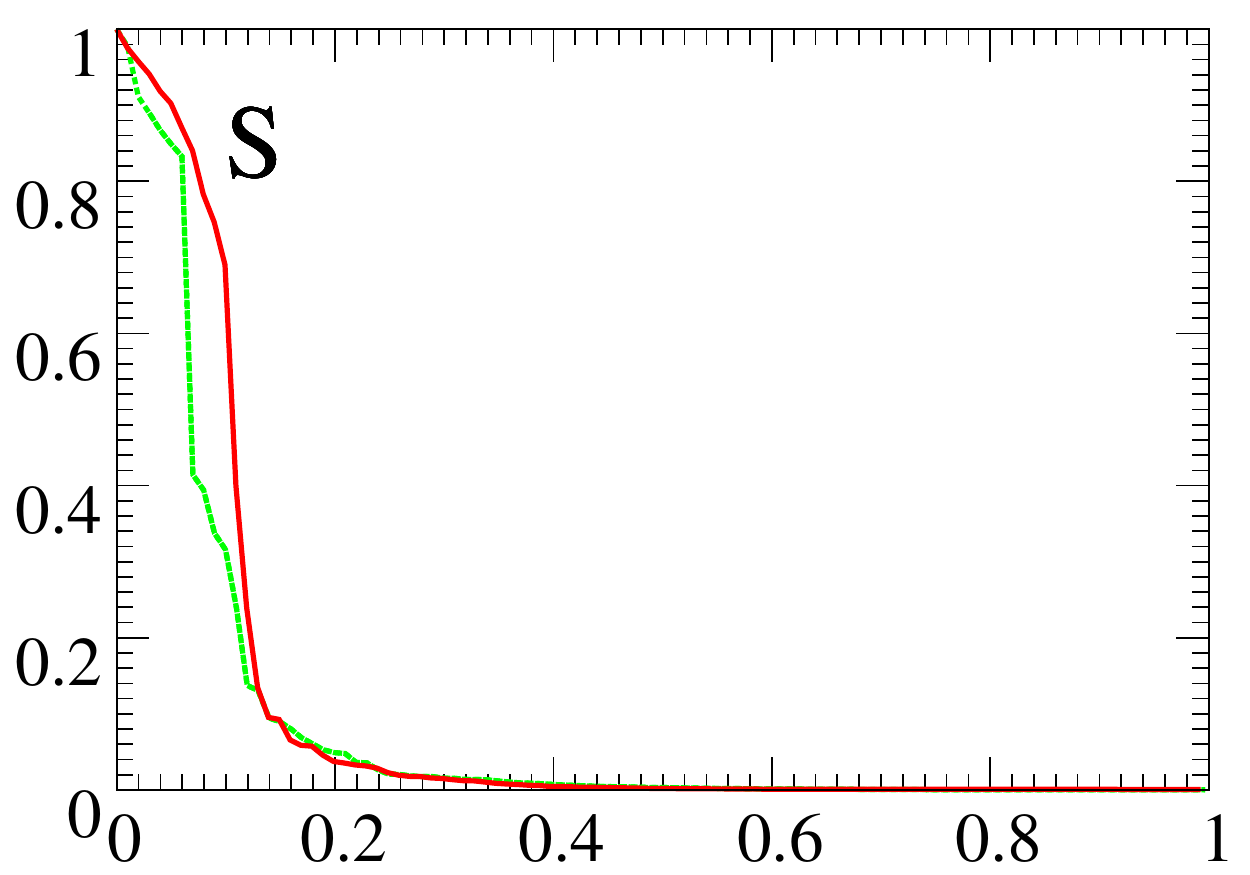}
    \includegraphics[width=0.32\textwidth]{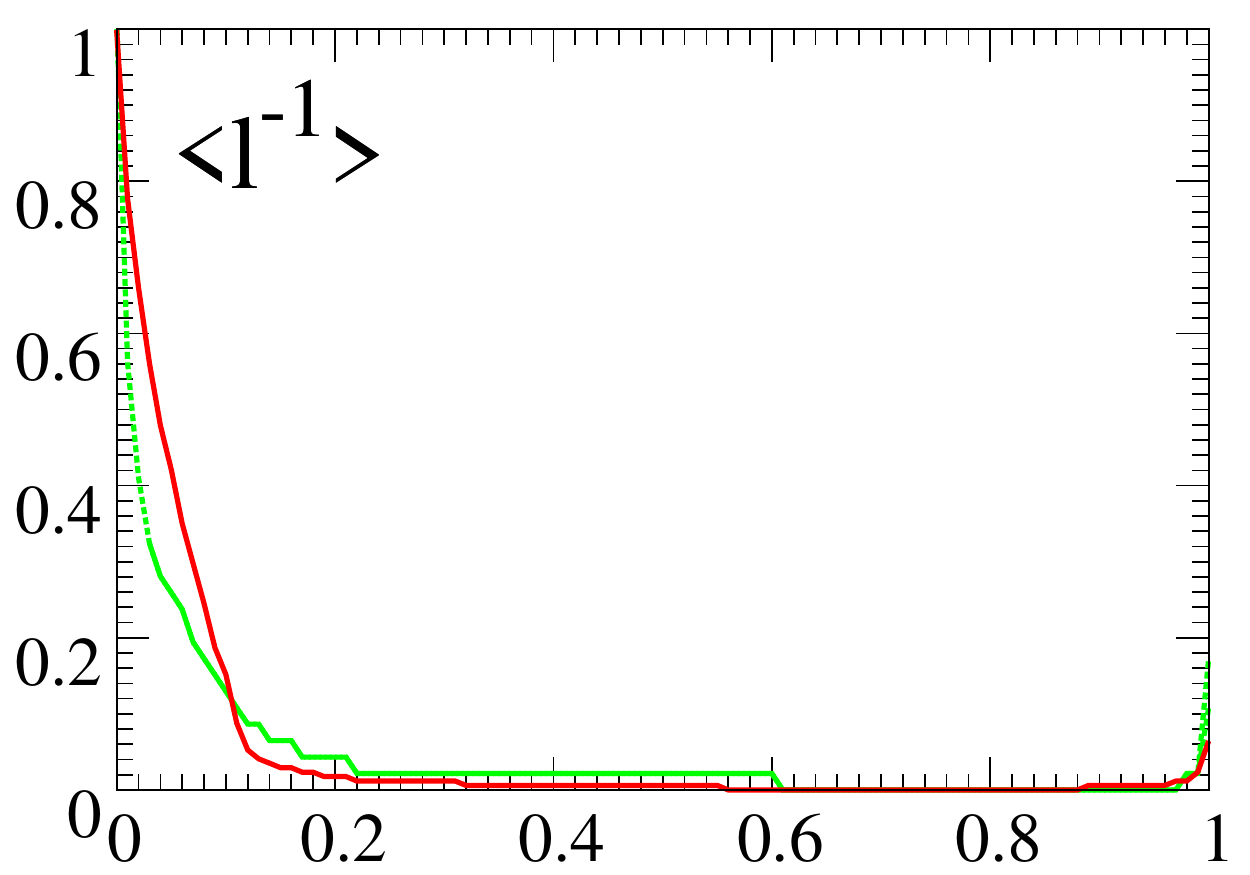}
\includegraphics[width=0.32\textwidth]{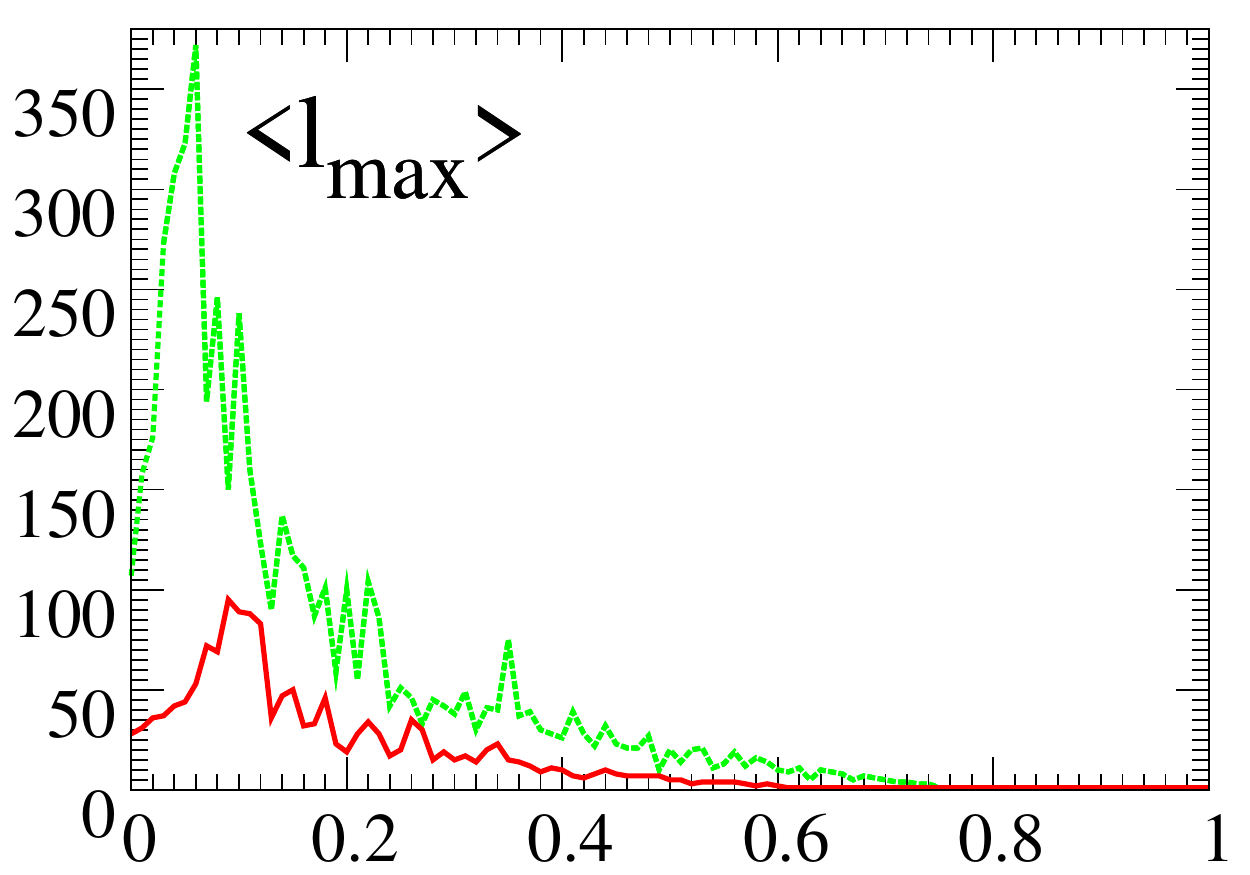}
\centerline{\bf d. \hspace{10em} e. \hspace{10em} f.}
\caption{\label{fig3}Share of the largest component $S$, mean
inverse $\langle \ell^{-1} \rangle$ and maximal $\ell^{\rm max}$
shortest path length as function of the removed nodes share $c$ for
PTN of London (light green curve) and Paris (dark red curve). {\bf
a}, {\bf b}, {\bf c}: highest node degree scenario. {\bf d}, {\bf
e}, {\bf f}: highest betweenness centrality scenario.}
\end{figure}

As clearly seen from Figs. \ref{fig2}, if nodes are removed ordered
by decreasing degree or betweenness centrality the size $S$ of the
largest component decreases fast and $S$ is near zero at a share of
removed nodes $c\sim 0.2 \div 0.3$. In Fig. \ref{fig3} we further
detail this picture giving plots for the size of the largest
component $S$, mean inverse $\langle \ell^{-1} \rangle$ and maximal
$\ell^{\rm max}$ shortest path lengths as function of the removed
node share $c$  for the highest node degree (figures {\bf a} -- {\bf
c}) and the highest betweenness centrality (figures {\bf d} -- {\bf
f}) scenarios. Let us compare these with the corresponding plots of
Fig. \ref{fig1}, where the impact of random node removal is shown.
We observe that for these directed attack scenarios the behavior of
both PTNs is not as different as it was observed for the random
scenario. Although for the highest node degree scenario both $S(c)$
and $\langle \ell^{-1} (c) \rangle$ curves manifest a faster decay
for the London PTN (see Figs. \ref{fig3} {\bf a}, {\bf b}), the
difference is less pronounced in the case of the highest betweenness
centrality scenario (Figs. \ref{fig3} {\bf d}, {\bf e}). The
similarity in the performance of both PTNs with respect to such
attacks is also obvious in the behavior of the  maximal shortest
path length $\ell^{\rm max}$. For both  London and Paris PTNs one
observes a pronounced peak in $\ell^{\rm max}(c)$  at $c\sim 0.06$
and $c\sim 0.1$  with further, smaller peaks occurring with
irregular periodicity indicating the existence of sub-clusters
within both networks.

The above comparison of the PTN attack vulnerability is as it stands
mostly qualitative. To proceed further with a quantitative analysis,
a numerical measure of resilience needs to be defined. In
percolation theory, where a spanning cluster occurs abruptly at a
given percolation concentration $c_{\rm perc}$, the latter may be
used as such a measure. In the case of real-world networks of finite
size one rather observes a region of concentrations where the
emergent behavior of fast decay occurs. In some studies a
characteristic concentration value based on particular behavior of
either  $S$, $\langle \ell \rangle$, $\langle \ell^{-1} \rangle$ or
$\ell^{\rm max}$ has been used to identify network breakdown
\cite{Holme02,attacks}. Here, we focus on the behavior of the
largest component of the PTN and follow Ref. \cite{Schneider} to
introduce a measure that integrates the network reaction over the
whole attack sequence. If $S(c)$ is the normalized size of the
largest component as function of concentration $c$, we calculate the
area $A$ below the $S(c)$ curve as:
\begin{equation}\label{11}
A = 100\int_0^1 S(c) {\rm d}c,
\end{equation}
and use this as a measure of network resilience. As  follows from
the definition (\ref{11}), the measure captures the effects on the
network over the complete attack sequence and it is a characteristic
measure, well-defined for finite-size networks. The larger the
measure $A$, the more resilient is the network.

In the left part of table \ref{tab2}  we give the resilience $A$ for
the highest node degree and highest betweenness scenarios and
compare with the random scenario. As follows from the table, in
almost all instances the Paris PTN shows higher resilience $A$ than
the London PTN.  Another conclusion concerns the difference between
the value of $A$ for the random attack (RV) and for attacks that
target specific important nodes (with high degree $k$ or high
betweenness centrality ${\cal C}_{B}$): as often observed for
complex networks, they are robust with respect to random removal of
nodes or links but especially vulnerable to targeted attacks.
Naturally the question arises whether such result may be anticipated
a priory: can one derive some criteria for PTN resilience prior to
the attack? Indeed, the data of Table \ref{tab1} where information
about initial PTN characteristics is summarized allows to at least
qualitatively prognosticate the outcome of attacks as summarized in
Table \ref{tab2}. For an explanation, let us shortly recall several
facts drawn from complex network theory.

\begin{table}
\caption{\label{tab2}Resilience measure $A$, (\ref{11}), for the
PTNs of London and Paris. Columns 2-4 give the value of $A$ for
node-targeted attacks, columns 5-7 give $A$ for link-targeted
attacks. See the text for attack scenario descriptions.}
\begin{center}
\tabcolsep1.7mm
 {\small
 \begin{tabular}{lrrrrrrrr}
 \hline\noalign{\smallskip}
 \multicolumn{1}{c}{City}& &
 \multicolumn{3}{c}{Node-targeted attacks}& &
 \multicolumn{3}{c}{Link-targeted attacks}\\
&\hspace{2em} & RV & $k$ & ${\cal C}_{B}$ & \hspace{2em} &RL&
$k^{(l)}$ &
 ${\cal C}_{B}^{(l)}$  \\
\noalign{\smallskip}\hline\noalign{\smallskip}
London       & &  29.31 &   5.45 &       8.71 &  &  27.45 &   20.95 &    27.2  \\
Paris        &  & 37.93 &  10.77 &     10.67 &   & 56.04 &   47.12 &   55.93 \\
\noalign{\smallskip}\hline
\end{tabular}
}
\end{center}
\end{table}

For uncorrelated infinite random networks it has been shown
\cite{Molloyreed,Cohen00}, that a giant connected component is
present if the following ratio of moments of the degree distribution
\begin{equation}\label{13}
\kappa^{(k)}= \langle k^2 \rangle/ \langle k \rangle,
\end{equation}
is greater than two,
\begin{equation}\label{14}
\kappa^{(k)}\geq 2.
\end{equation}
Relation (\ref{14}) is often referred to as the Molloy-Reed
criterion and $\kappa^{(k)}$ is called the Molloy-Reed parameter.

It has been illustrated for many real-world PTN
\cite{Holovatch11,attacks,Berche12}, that the value of the
Molloy-Reed parameter for the unperturbed network may be used to
estimate network resilience against attack. Typically, networks with
low $\kappa^{(k)}$ appear to be more vulnerable to both random and
node degree-targeted attacks. This observation is further supported
by monitoring other related parameters, such as the ratio of the
mean number $z_2$ of second neighbors to the mean number $z_1$ of
neighbors:\footnote{By definition  $z_1$ is equal to the mean node
degree $\langle k \rangle$.}
\begin{equation}\label{15}
\kappa^{(z)}=z_2/z_1.
\end{equation}
It is easy show that $\kappa^{(k)}=\kappa^{(z)}+1$ for uncorrelated
networks. As we have seen from the analysis of section \ref{II},
strong correlations are present in the PTN and one may not expect a
simple relation between parameters $\kappa^{(k)}$  and
$\kappa^{(z)}$  to hold. However, a comparison of $\kappa^{(z)}$ for
two given networks will provide additional information about their
relative resilience.

We have calculated values of $\kappa^{(k)}$ and $\kappa^{(z)}$ for
the London and Paris PTNs and give them in the ninth and tenth
columns  of table \ref{tab1}. The corresponding values for the Paris
PTN exceed those for London  by a factor of two giving a clear
signal for a higher vulnerability of the London PTN to random
failure. This conclusion has been empirically demonstrated in our
simulations reported above.

The higher potential for resilient behavior of the  Paris PTN with
respect to that of London may also be related to its node-degree
distribution. In the last column of table \ref{tab1} we list the
exponent $\gamma$, that controls the decay of this distribution. The
smaller value of $\gamma$ for Paris PTN corresponds to the
fat-tailed node-degree distribution $P(k)$. For an infinite network,
the giant connected component is always present for the random
attack scenario as long as $\gamma<3$ \cite{Cohen00} and a smaller
values of $\gamma$ indicate higher network resilience.

Our analysis has so far described attacks  on the network nodes.
Before passing to general conclusions, let us further analyze the
impact of link-targeted attacks on the two PTNs.

\subsection{Link-targeted attacks} \label{IV.2}

Considering link-targeted attacks we concentrate here on those
scenarios that have proven to be most harmful for node-targeted
variants, namely removing links with highest degree and highest
betweenness centrality. Our aim is to check how resilient the two
PTNs are to attacks on links following the corresponding scenarios.
However, to proceed we need to generalize the notions of degree and
betweenness for the case of links. We define the degree $k^{(l)}$ of
the link between nodes $i$ and $j$ with degrees $k_i$ and $k_j$ as
\cite{Holovatch11,Berche12}:
\begin{equation}\label{16}
k^{(l)}_{ij} = k_i + k_j - 2.
\end{equation}
With this definition, the link degree is $k^{(l)}=0$ for a graph
with two vertices and a single link, while for any link in a
connected graph with more than two vertices the link degree will be
at least one, $k^{(l)} \geq 1$. The generalization of betweenness
centrality for a link $e$ is straightforward:
\begin{equation}\label{17}
{\cal C}_{B}^{(l)}(e)= \sum_{s\neq t\in \cal{N}}
\frac{\sigma_{st}(e)}{\sigma_{st}},
\end{equation}
where $\sigma_{st}$ is the number of shortest paths between the two
nodes $s,t\in \cal{N}$, that belong to the network $\cal{N}$, and
$\sigma_{st}(e)$ is the number of shortest paths between nodes $s$
and $t$ that go through the link $e$ (c.f. formula (\ref{5}) for the node
betweenness centrality).  By definition, ${\cal C}_{B}^{(l)}(e)$
measures the importance of a link $e$ with respect to the
connectivity between the nodes of the network.

\begin{figure}[h]
\includegraphics[width=0.75\textwidth]{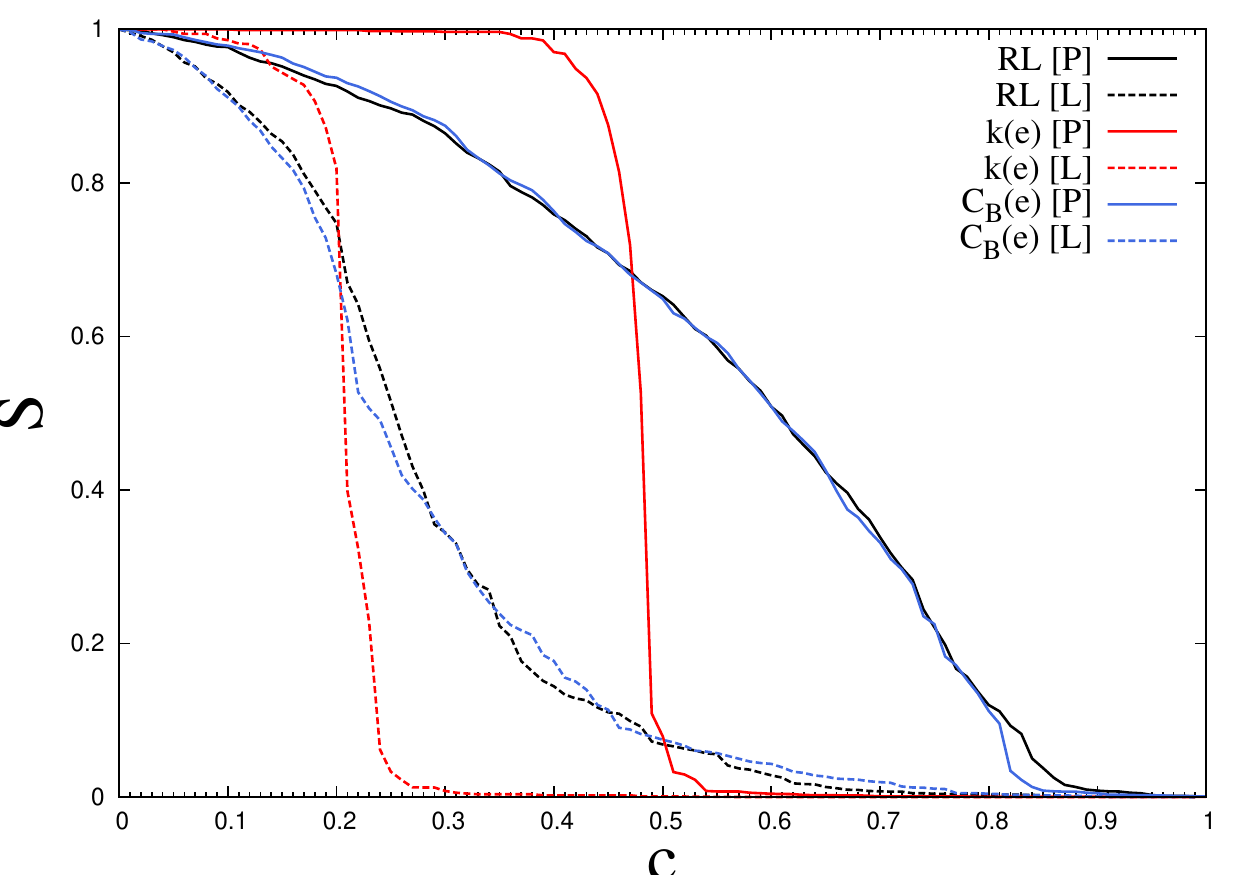}
\caption{Relative size of the largest component of the Paris and
London PTNs for different link attack scenarios. RL: random removal
of a link, $k(e)$, ${\cal C}_B(e)$: highest link degree and highest
link betweenness scenarios.  Here, the links (not the nodes) are
removed. Hence, $c$ denotes the share of removed links. As in Fig.
\ref{fig2}, a letter in square brackets refer to the London [L] or
Paris [P] PTN.} \label{fig4}
\end{figure}

Fig. \ref{fig4} shows the results of our simulations for three
different attack scenarios, where the PTN links are removed at
random (RL) or according to lists ordered by decreasing link degrees
and link betweenness centrality. As in the case of node-targeted
attacks these lists were recalculated after each step of 1\% of link
removal. The figure shows the relative size of the largest component
of the PTN as function of the share of removed links. Let us first
note that the removal of a link does  not necessarily lead to a
decrease in $S$. Indeed, as we see from the figure $S$ may remain
unchanged for small enough values of $c$, depending on the attack
scenario. This is different from the node-targeted attacks, where
the removal of a node decreases the size of $S$ at least by the
relative share of this node. In this respect, the most particular
behavior is observed for the highest link degree scenario (red
curves in Fig. \ref{fig4}). The value of $S$ first remains
practically unchanged (up to a concentration of removed links $c\sim
0.08$ for London PTN and even $c\sim 0.36$ for Paris PTN) and then
abruptly decreases almost to zero. This behavior however is an
artifact of the recalculated link degree scenario: after removal of
the top 1\% of links linked to highest degree nodes these nodes may
remain connected and will in general not be targeted in the next
step after recalculation. Therefore many steps are needed to strip
these nodes off all their links. To further quantify the impact of
different scenarios we have calculated the value of the resilience
measure $A$, introduced in the previous section, see Eq. (\ref{11}).
We present the results for all three scenarios in Table \ref{tab2}.
As shown in the table, almost for all link-targeted scenarios the
values of $A$ is almost twice as large for the Paris PTN in
comparison with the London PTN. Another obvious observation is that
different scenarios applied to the same PTN lead to similar values
of $A$. Returning back to Fig. \ref{fig4} it is obvious that not
only the resilience measure $A$ but also the $S(c)$ curves
demonstrate very similar behavior following both random and link
betweenness scenarios.

 Based on the above simulated attack scenarios
we observe that under almost all of these the London PTN appears to
be more vulnerable than the  Paris PTN. One may therefore ask if
there are a-priory criteria that may indicate network resilience
prior to any (simulated) attack.
 In former analysis \cite{Holovatch11,Berche12} we found, that a useful
criterion for resilience of PTNs with respect to link-targeted
attacks is the mean node degree $\langle k \rangle$ of the
unperturbed network. Typically, networks with a higher mean node
degree are more resilient. Furthermore, in a recent study on the
link-targeted resilience of fourteen different PTNs of major cities
\cite{Holovatch11,Berche12}, the resilience measure $A$ was found to
almost  linearly increase with $\langle k \rangle$. This appears to
indicate that network (link) resilience depends primarily on the
initial 'density' of network  links, almost independent of possible
correlations within the PTN structure. To some extent this is
different to the criteria discussed in the former subsection for the
node-targeted attacks, where the correlations were considered
involving the second moment of the node degree distribution $\langle
k^2 \rangle$ that enters the Molloy-Reed parameters (\ref{13}),
(\ref{15}). Comparing $\langle k \rangle$ for the two unperturbed
PTNs (table \ref{tab1}) one can see that its value for Paris PTN
exceeds that for London PTN almost in 1.4 times (2.60 for London and
3.73 for Paris, see the table). This observation may be taken as an
indicator for a correspondingly higher resilience of the Paris PTN.

\begin{figure}[h]
\includegraphics[width=0.28\textwidth]{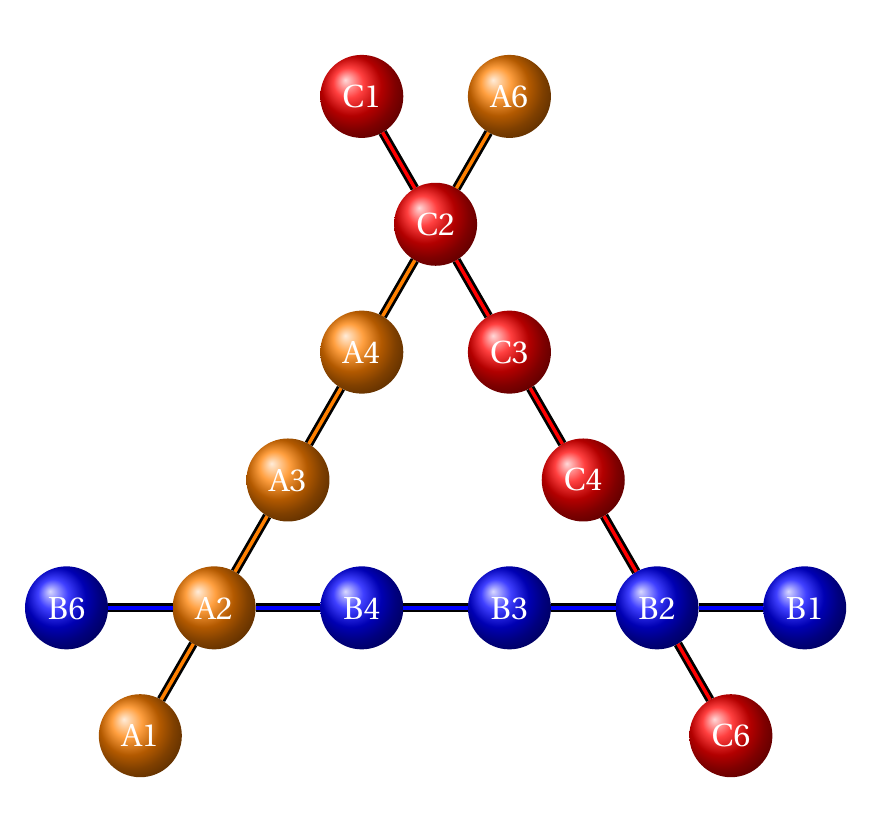}\hfill
\includegraphics[width=0.28\textwidth]{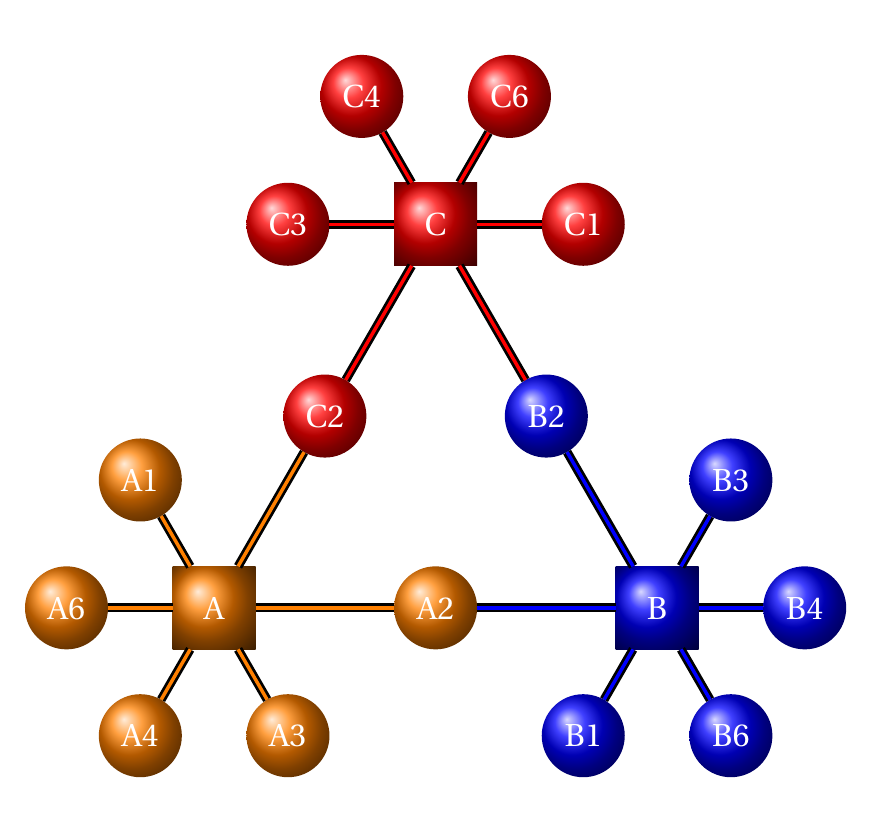}\hfill
\includegraphics[width=0.28\textwidth]{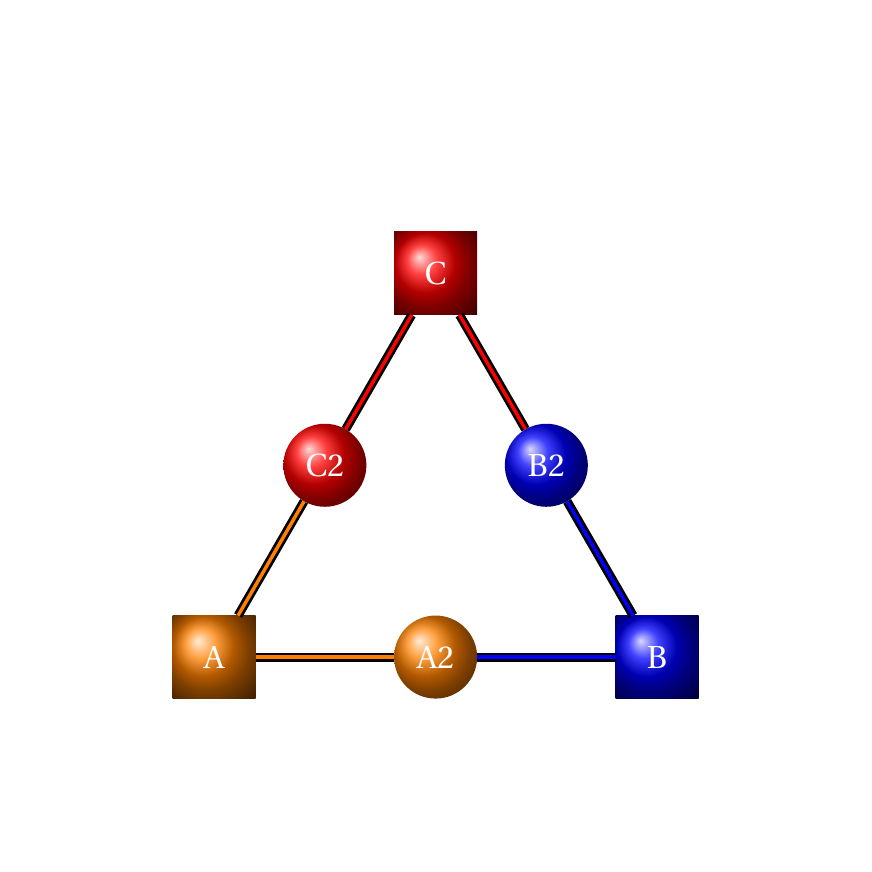}\hfill\\
\hspace*{\fill} {\bf a.} \hfill\hfill {\bf b.} \hfill\hfill {\bf c.}
\hspace*{\fill} \caption{\label{fig5} a: A graph with three routes,
each shown in a separate color; b: the corresponding bipartite graph
- route nodes are depicted as square boxes; c: the weeded graph
without dangling station nodes.}
\end{figure}

\section{One step further: Cascading effects}
\label{V} The approaches to network attack as described in the
previous sections  assume that any attack on a given station will in
first place affect the attacked station and the links to its direct
neighbors within the network. The operation of all traffic on the
transit network is in this view unaffected on all other links and
nodes within the network. This implies some non-realistic
situations: e.g. if a subway station X on the London tube ceases to
function the model assumes that all routes that would otherwise pass
through that station will be split into two halves that continue to
function as normal on the remaining parts of that route. Obviously
this will in general not happen and instead the route as a whole
will cease to operate or at least be seriously reduced in its
function.

\begin{figure}[h]
\includegraphics[width=1\textwidth]{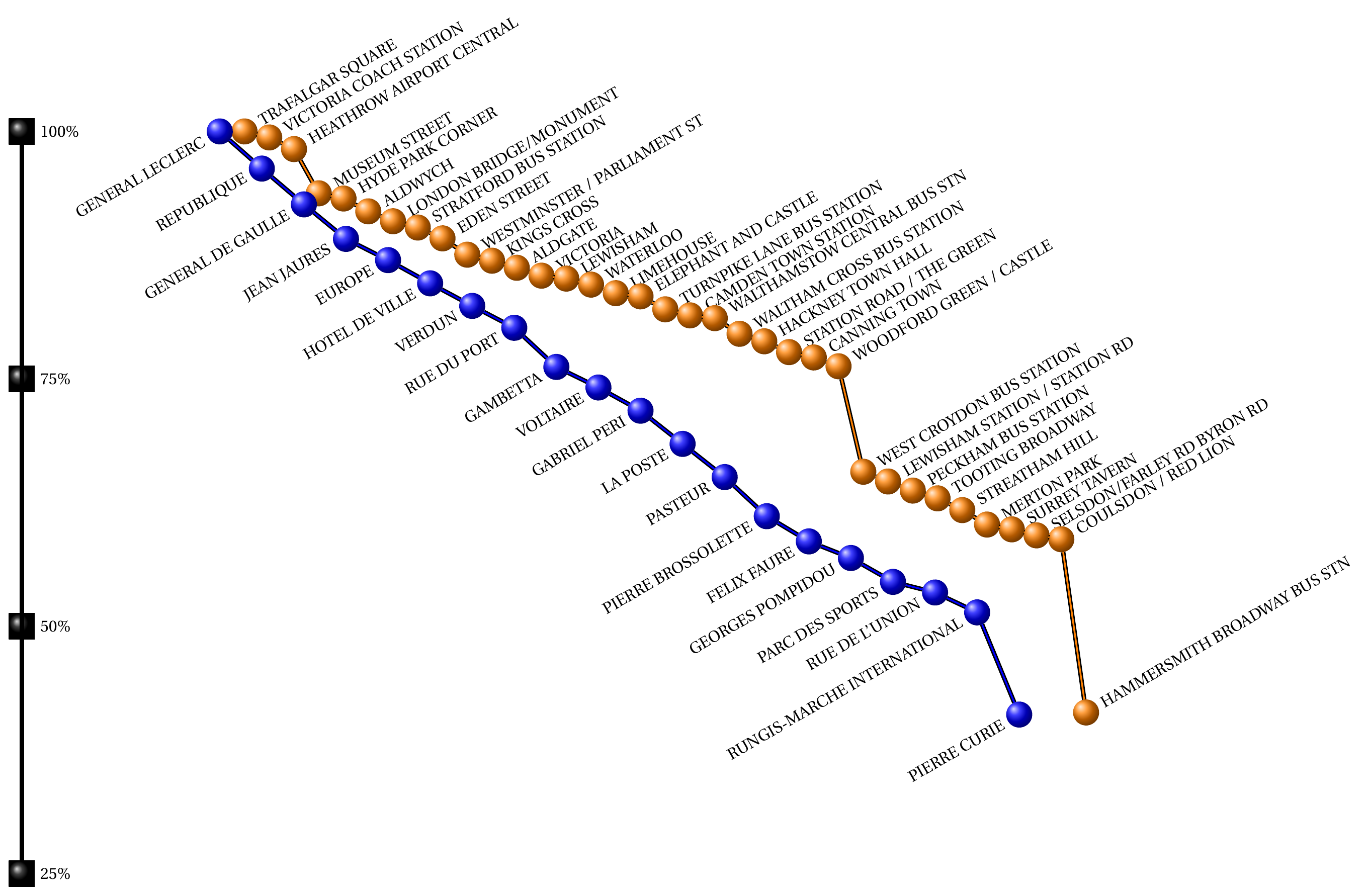}
\caption{\label{fig6} The break down of the connected component of
the London (light yellow) and Paris (dark blue) PTN under cascading
effects, see the text for the attack scenario. For each step of the
attack we depict the corresponding station with highest betweenness
that is subsequently removed. The axis on the left indicates the
remaining percentage of the connected part of the network.}
\end{figure}

We therefore embark in this final section to explore the impact of
attacks on the network including the cascading effects on all routes
that service the attacked station in assuming that all routes that
service that station will cease to operate.

This task may be considerably simplified by re-interpreting the
network in terms of a so-called bipartite graph. The procedure is
illustrated in Fig. \ref{fig5}.  Every route is represented by a
square vertex connected to the nodes of all its stations, see
\ref{fig5}a,b. In a further simplifying step we weed out all
stations that are connected to a single route only, as they do not
contribute to the connectivity of the network, see \ref{fig5}c.

Within this bipartite network we identify the station node with
highest betweenness following the same procedure as above. That node
and all adjacent routes are then removed - as will all station nodes
that become disconnected in this process.

The latter step is repeated until the largest connected component in
the remaining network is smaller than half of the original bipartite
graph indicating complete breakdown of the network.

Following this procedure we find that both the Paris and the London
network reach the 50\% breakdown point when only 0.47\% of the total
stations become dysfunctional. This corresponds to 34 stations of
the London PTN and 19 stations of the Paris network. Fig. \ref{fig6}
depicts the break down of the connected component of the network.
For each step of the procedure we depict the corresponding station
with highest betweenness that is subsequently removed. The axis on
the left indicates the remaining percentage of the connected part of
the network. We close by noting that  on an operational basis the
network may break down even much earlier, than predicted by our
theory as far as the load to be transferred to the remaining routes
will far exceed by far the capacity of these at an even lower number
of dysfunctional routes.

\begin{acknowledgements}
We thank Ralph Kenna for helpful discussions and a suggestion for
the title. This work was supported by FP7 grant SPIDER (Statistical
Physics in Diverse Realisations) PIRSES-GA-2011-295302.
\end{acknowledgements}



\end{document}